%% file: block.tex
\documentclass[conference]{IEEEtran}

\input{preamble}

\usepackage{cite}
\usepackage{printlen}

\title{Block-to-Block Distribution Matching}

\IEEEoverridecommandlockouts

\author{\IEEEauthorblockN{Georg B\"ocherer, Rana Ali Amjad}
\IEEEauthorblockA{Institute for Communications Engineering\\Technische Universit\"at M\"unchen, Germany\\
Email: \texttt{georg.boecherer@tum.de,raa2463@gmail.com}}
\thanks{This work was supported by the German Ministry of Education and Research in the framework of an Alexander von Humboldt Professorship.}
}

\DeclareMathOperator{\supp}{supp}

\begin{document}

\maketitle

\begin{abstract}
In this work, binary block-to-block distribution matching is considered. $m$ independent and uniformly distributed bits are mapped to $n$ output bits resembling a target product distribution. A rate $R$ is called achieved by a sequence of encoder-decoder pairs, if for $m,n\to\infty$, (1) $m/n\to R$, (2) the informational divergence per bit of the output distribution and the target distribution goes to zero, and (3) the probability of erroneous decoding goes to zero. It is shown that the maximum achievable rate is equal to the entropy of the target distribution. A practical encoder-decoder pair is constructed that provably achieves the maximum rate in the limit. Numerical results illustrate that the suggested system operates close to the limits with reasonable complexity. The key idea is to internally use a fixed-to-variable length matcher and to compensate underflow by random mapping and to cast an error when overflow occurs.
\end{abstract}
\section{Introduction}
Binary distribution matching refers to reversibly mapping independent and uniformly distributed bits to bits that are distributed approximately according to a target distribution. The degree of approximation is measured by the informational divergence (I-divergence). In digital communication systems, distribution matchers can be used, e.g., for coding for noiseless channels  or for probabilistic shaping for noisy channels \cite[Chap. 6 \& 7]{bocherer2012capacity}. \emph{Variable-to-fixed length} (v2f) matchers are developed in \cite[Chap.~3]{bocherer2012capacity} and the  authors of this paper propose \emph{fixed-to-variable length} (f2v) matchers in \cite{amjad2013fixed}. 

Variable length matchers have three inherent problems, namely \emph{synchronization}, \emph{error propagation}, and \emph{variable transmission rate}. We illustrate this by an example. Consider the v2f matcher
\begin{align}
1\mapsto a, 00\mapsto b, 01\mapsto c
\end{align}
which generates the channel input symbols $\{a,b,c\}$ with probabilities $1/2,1/4,1/4$, respectively. The binary string $01001$ is mapped to $cba$, which is then transmitted over a noisy channel. The string $aba$ is detected at the receiver and according to the matcher mapped to $1001$, i.e.,
\begin{align*}
&01001\mapsto cba\\
\textcolor{red}{a}ba\mapsto &1001.
\end{align*}
First, the input length is $5$ but the output length is $4$, so input and output are out of sync. Second, one detection error led to 3 bit errors and one bit is missing. Third, an all $b$ string on the channel corresponds to twice as many data bits then an all $a$ string of the same length. Thus, a system that deploys a variable length matcher needs the capability to buffer large amounts of data to keep up with the variable transmission rate. 
\begin{figure}[t]
\centering
\footnotesize
\def\svgwidth{0.9\columnwidth}
\executeiffilenewer{figures/fixedVsVariable.svg}{figures/fixedVsVariable.pdf}%
{inkscape -z -D --file=figures/fixedVsVariable.svg %
--export-pdf=figures/fixedVsVariable.pdf --export-latex}%
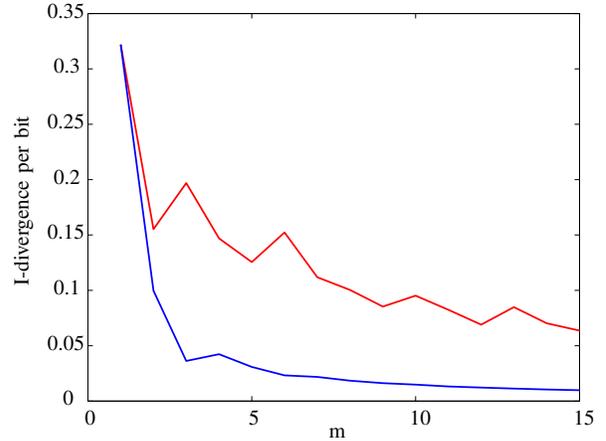%

\caption{Comparison of optimal one-to-one b2b matching (red line) and optimal fixed-to-variable length matching (blue line) for the target distribution $P_Y(0)=1-P_Y(1)=0.8$. The horizontal axis displays the number of input bits and the vertical axis displays the I-divergence per bit. }
\label{fig:oto}
\vspace{-0.5cm}
\end{figure}

The three drawbacks of variable length matchers stated above motivate us to investigate the design of block-to-block (b2b) matchers that map $m$ input bits to $n$ output bits. The ratio $m/n$ is called the \emph{matching rate}. For b2b matchers, the transmission rate is constant and synchronization errors and error propagation are limited by the block length.

In theory, optimal one-to-one b2b matchers can easily be constructed, we detail this in Subsec.~\ref{subsec:optimal}. However, these matchers may not be practical. We illustrate this by an example. For the target distribution $P_Y(0)=1-P_Y(1)=0.8$, the I-divergence per bit versus the input length $m$ that is achieved by an optimal one-to-one b2b matcher is plotted in Fig.~\ref{fig:oto} by a red curve. For comparison, the performance of an optimal v2f matcher is plotted by a blue curve. For the same codebook size, the v2f matcher achieves an I-divergence that is around five times smaller than the I-divergence achieved by the b2b matcher. Basically, Fig.~\ref{fig:oto} suggests that b2b matchers cannot achieve a low I-divergence with a reasonable complexity. 

In this work, we show how b2b matchers can be constructed that provably achieve the same I-divergence as the f2v matcher with the same complexity. The key idea is to repeatedly use the f2v matcher inside the b2b matcher. For a fixed output length, this results in underflow events and overflow events. We handle underflow events by random mapping and overflow events by casting an error. The probability of error can be made arbitrarily small by choosing the block size large enough. This corresponds to increasing the number of times the v2f matcher is applied internally and it does not effect the complexity of the b2b matcher. We call b2b matcher with a small probability of decoding error \emph{$\epsilon$-error b2b matcher}.

In Sec.~\ref{sec:problem} we precisely define the $\epsilon$-error b2b design problem. We then show in Sec.~\ref{sec:converse} that no matching rate larger than the entropy of the target distribution can be achieved. In Sec.~\ref{sec:noerror}, we address (impractical) zero-error b2b matcher. The construction of a (practical) $\epsilon$-error b2b matcher is described in Sec.~\ref{sec:error} and its asymptotic optimality is proven in Sec.~\ref{sec:analysis}. We finally give numerical results in Sec.~\ref{sec:numerical} that illustrate that the suggested $\epsilon$-error b2b matcher has the same performance as the f2v matcher with a small probability of error.
\section{Problem Statement}
\label{sec:problem}
\subsection{Matching}
Let $B^m$ be a sequence of $m$ binary random variables that are independent and uniformly distributed. An \emph{encoder} is a mapping
\begin{align}
f_n\colon\{0,1\}^m\to\{0,1\}^n,\quad B^m\mapsto f_n(B^m)=:\tilde{Y}^n.
\end{align}
We allow $f_n$ to be a random mapping. The corresponding \emph{decoder} is a mapping
\begin{align}
\varphi_n\colon\{0,1\}^n\to\{0,1\}^m,\quad \tilde{Y}^n\mapsto \varphi_n(\tilde{Y}^n)=:\hat{B}^m.
\end{align}
Let $P_Y$ be a binary target distribution with $0<P_Y(0)<1$ and $P_Y(1)=1-P_Y(0)$. Denote by $P_Y^n$ the joint distribution of $n$ binary random variables that are iid according to $P_Y$. For a given $P_Y$, we say that a \emph{matching rate} $R$ is \emph{achievable} if there exists a sequence of encoder-decoder pairs $\{f_n,\varphi_n\}_{n=1}^\infty$ that fulfills the following three conditions as \text{$n\to\infty$}.
\begin{align}
\frac{m}{n}\to &R\label{eq:rate}\\
\frac{\kl(P_{\tilde{Y}^n}\Vert P_Y^n)}{n}\to &0\label{eq:divergence}\\
\probop(B^m\neq\hat{B}^m)\to &0.\label{eq:error}
\end{align}
Note that $m$ implicitly depends on $n$.
\subsection{Maximum Matching Rate}

We illustrate by an example that the rate $R=0$ can easily be achieved.

\emph{Example:} Define $f_n$ as
\begin{align}
b&\mapsto f_n(b)=b\tilde{Y}^{n-1},\quad b\in\{0,1\}
\end{align}
where $\tilde{Y}^{n-1}$ is distributed according to $P_Y^{n-1}$.
By the chain rule, the informational divergence is given by
\begin{align}
&\kl(P_{\tilde{Y}^n}\Vert P_Y^n)\nonumber\\
&=\kl(P_{\tilde{Y}_1}\Vert P_Y)+\sum_{y\in\{0,1\}}P_{\tilde{Y}_1}(y)\kl(P_{\tilde{Y}_2^n|\tilde{Y}_1=y}\Vert P_{Y_2^n|Y_1=y})\nonumber\\
&=\kl(P_{\tilde{Y}_1}\Vert P_Y).
\end{align}
Thus, as $n\to\infty$, \eqref{eq:divergence} is fulfilled. We define the decoder as
\begin{align}
\tilde{y}^n\mapsto\varphi_n(\tilde{y}^n)=\tilde{y}_1.
\end{align}
Clearly, it decodes with an error probability of zero and \eqref{eq:error} is fulfilled. Thus, $\{f_n,\varphi_n\}$ fulfills our requirements for a matcher with a matching rate of $1/n\to0$. 

This example shows that ``small'' rates can easily be achieved by using random mappings from input symbols to disjoint sets of output symbols. Such mappings allow error free decoding. We are therefore interested in the \emph{maximum}  achievable matching rate.

\section{Converse}
\label{sec:converse}
We will need the following implication, which is shown in \cite{bocherer2013rooted}. Convergence in normalized informational divergence implies convergence in entropy rate, i.e., as $n\to\infty$,
\begin{align}
\frac{\kl(P_{\tilde{Y}^n}\Vert P_Y^n)}{n}\to 0\Rightarrow \frac{\entop(P_{\tilde{Y}^n})}{n}\to\entop(P_Y).\label{lem:entropy}
\end{align}

\begin{proposition}\label{prop:converse}
Let $\{f_n,\varphi_n\}_n$ be a sequence of encoder-decoder pairs that for a target distribution $P_Y$ achieves a matching rate of $R$. Then
\begin{align}
R\leq\entop(P_Y).
\end{align}
\end{proposition}
\begin{IEEEproof}
\emph{Estimating $B^m$ from $\tilde{Y}^n$:} Denote by $\hat{B}^m$ an estimate of $B^m$ that results from processing $\tilde{Y}^n$. Define $P_e:=\probop(B^m\neq\hat{B}^m)$ and $\entop_2(p):=-p\log_2p-(1-p)\log_2(1-p)$. Then, by Fano's inequality \cite[Sec. 1.9.2]{kramer2012information} we have
\begin{align}
\entop_2(P_e)+&P_e\log_2(|\{0,1\}|^m)=\entop_2(P_e)+P_em\\
&\geq\entop(B^m|\hat{B}^m)\\
&=\entop(B^m)-[\entop(B^m)-\entop(B^m|\hat{B}^m)]\\
&=m-\miop(B^m;\hat{B}^m)\\
&\overset{(a)}{\geq} m-\miop(B^m;\tilde{Y}^n)\\
&=m-\entop(\tilde{Y}^n)+\entop(\tilde{Y}^n|B^m)\\
&\geq m-\entop(\tilde{Y}^n)
\end{align}
where we used the data processing inequality \cite[Theo.~1.4]{kramer2012information} in (a).
Dividing by $m$, we get
\begin{align}
\frac{\entop_2(P_e)}{m}+P_e\geq 1-\frac{\entop(\tilde{Y}^n)}{m}.
\end{align}
By \eqref{lem:entropy}, we know that $\entop(\tilde{Y}^n)/n\to \entop(P_Y)$. Thus, in the limit,
\begin{align}
\frac{\entop_2(P_e)}{m}+P_e\geq 1-\frac{n}{m}\entop(P_Y)
\end{align}
Thus, if the rate $m/n$ is larger than $\entop(P_Y)$, then the probability of error is bounded away from zero. This is the statement of the proposition.
\end{IEEEproof}

\section{Zero-Error Achievability}
\label{sec:noerror}

\begin{proposition}
For any binary target distribution $P_Y$, the maximum matching rate $\entop(P_Y)$ can be achieved by a zero-error matcher.
\end{proposition}
\begin{IEEEproof} The proof is based on typicality. Fix $\epsilon>0$. Denote by $T_\epsilon^n(P_Y)$ the set of length-$n$ sequences that are $\epsilon$-letter typical with respect to $P_Y$. By \cite[Theorem~4.2]{kramer2012information}, the cardinality of this set is lower bounded by
\begin{align}
|T_\epsilon^n(P_Y)|\geq [1-\delta_\epsilon(P_Y,n)]2^{n(1-\epsilon)\entop(P_Y)}
\end{align}
where $\delta_\epsilon(P_Y,n)\overset{n\to\infty}{\longrightarrow}0$. In particular, there exists an $n_0$, such that for all $n\geq n_0$, $\delta_\epsilon(P_Y,n)\leq\frac{1}{2}$. Assume $n\geq n_0$. We choose $m$ such that there are $2^m$ distinct typical sequences:
\begin{align}
m&=\Bigl\lfloor \log_2[1-\delta_\epsilon(P_Y,n)]\Bigr\rfloor+\Bigl\lfloor n(1-\epsilon)\entop(P_Y)\Bigr\rfloor\label{eq:mchoice}\\
&\geq -1 + n(1-\epsilon)\entop(P_Y)-1\\
&=n(1-\epsilon)\entop(P_Y)-2.\label{eq:mbound}
\end{align}
Let $\mathcal{C}\subseteq T_\epsilon^n(P_Y)$ be a set of $2^m$ typical sequences. We define the encoder $f_n$ as a one-to-one mapping from $\{0,1\}^m$ to $\mathcal{C}$ and we define the decoder as $\varphi_n=f_n^{-1}$. We now verify conditions \eqref{eq:rate}--\eqref{eq:error}.

\emph{Probability of error:} Since the defined mapping is one-to-one, the probability of error is equal to zero for any $n$.

\emph{I-Divergence:} 
For each sequence $y^n\in\mathcal{C}$, the probability $P_Y^n(y^n)$ is by \cite[Theorem~4.2]{kramer2012information} bounded as
\begin{align}
P_Y^n(y^n)\geq 2^{-n(1+\epsilon)\entop(P_Y)}\label{eq:pbound}.
\end{align}
We calculate 
\begin{align}
\kl(P_{\tilde{Y}^n}\Vert P^n_Y)=&\sum_{y^n\in\mathcal{C}}2^{-m}\log_2\frac{2^{-m}}{P_Y^n(y^n)}\\
\overset{\text{(a)}}{\leq}&\sum_{y^n\in\mathcal{C}}2^{-m}\log_2\frac{2^{-m}}{2^{-n(1+\epsilon)\entop(P_Y)}}\\
 =&n(1+\epsilon)\entop(P_Y) - m\\
\overset{\text{(b)}}{\leq} &2\epsilon n\entop(P_Y)+2
\end{align}
where (a) follows from \eqref{eq:pbound} and where (b) follows from \eqref{eq:mbound}.
Thus,
\begin{align}
\lim_{n\to\infty}\frac{\kl(P_{\tilde{Y}^n}\Vert P^n_Y)}{n}\leq&\lim_{n\to\infty}\frac{2\epsilon n\entop(P_Y)+2}{n}\\
=&2\epsilon \entop(P_Y).\label{eq:exdivergence}
\end{align}
This holds for any $\epsilon>0$, which shows that condition \eqref{eq:divergence} is fulfilled.

\emph{Rate:} By \eqref{eq:mbound}, the rate $m/n$ is bounded as
\begin{align}
(1-\epsilon)\entop(P_Y)-\frac{2}{n}\leq \frac{m}{n}.\label{eq:exentropy}
\end{align}
Thus, as $\epsilon\to 0$ and $n\to\infty$, $m/n\geq \entop(P_Y)$.
\end{IEEEproof}
\emph{Remark:} Note that $\epsilon\to 0$ drives both the normalized informational divergence in \eqref{eq:exdivergence} to zero and the entropy rate in \eqref{eq:exentropy} to $\entop(P_Y)$. This exemplifies the relation between informational divergence and entropy that we stated in \eqref{lem:entropy}.

\subsection{Optimal One-to-One b2b Matching}
\label{subsec:optimal}
The optimal one-to-one b2b matcher for a fixed input length $m$ chooses the output length $n$ and the codebook $\mathcal{C}\subseteq\{0,1\}^n$ for which the I-divergence per output bit is minimal. To find the optimal one-to-one b2b matcher, we need to solve the optimization problem
\begin{align}
\minimize_n\left\{\minimize_{\mathcal{C}\subseteq\{0,1\}^n\colon |\mathcal{C}|=2^m}\Bigl\{\frac{1}{n}\sum_{y^n\in\mathcal{C}}2^{-m}\log_2\frac{2^{-m}}{P_Y^n(y^n)}\Bigr\}\right\}.\nonumber
\end{align}
Minimizing over $n$ can be done by a line search around $n\approx m/\entop(P_Y)$ and for each $n$, the codebook $\mathcal{C}$ that minimizes the I-divergence is the one that contains the $2^m$ sequences from $\{0,1\}^n$ that are most probable according to $P_Y^n$.
\section{$\epsilon$-Error b2b Matching: Code Construction}
\label{sec:error}
\subsection{Fixed-to-Variable Length Matching \cite{amjad2013fixed}}
A fixed-to-variable length code is a mapping
\begin{align}
h\colon\{0,1\}^j\to\{0,1\}^+
\end{align}
where $^+$ denotes the \emph{Kleene plus}, i.e., $\{0,1\}^+$ is the set of all binary strings that have a length larger or equal to one. Suppose $B^j=b^j$. Denote by $\ell(b^j)$ the length of the code word $h(b^j)$. We denote the image of $h$ by
\begin{align}
h(\{0,1\}^j)=:\mathcal{C}\subseteq \{0,1\}^+.
\end{align}
The mapping $h$ defines a random variable $U$ that is uniformly distributed over $\mathcal{C}$, i.e,
\begin{align}
P_U(c)=2^{-j},\quad\forall c\in\mathcal{C}.
\end{align}
Also, we define a target distribution $P_Y^+$ in the following way.
\begin{align}
P_Y^+(c)=P_Y^{\ell(c)}(c).
\end{align}
Suppose $h$ is the optimal fixed-to-variable length code with respect to $P_Y$. Then, by \cite[Prop.~6]{amjad2013fixed}
\begin{align}
\frac{\kl(P_U\Vert P_Y^+)}{\expop[\ell(U)]}\overset{j\to\infty}{\longrightarrow}0.\label{eq:divergencef2v}
\end{align}
In the following, we will use this mapping $k$ times, i.e., we encode blocks of $m=kj$ binary symbols. We define 
\begin{align}
h^k\colon \{0,1\}^m\to&\mathcal{C}^k\\b^m\mapsto &h(b_1^j)h(b_{j+1}^{2j})\dotsb h(b_{(k-1)j+1}^{kj}).
\end{align}
Because of
\begin{align}
\frac{\kl(P_{U^k}\Vert P_Y^+)}{\expop[\ell(U^k)]}=\frac{k\kl(P_U\Vert P_Y^+)}{k\expop[\ell(U)]}
\end{align}
the limit \eqref{eq:divergencef2v} implies
\begin{align}
\frac{\kl(P_{U^k}\Vert P_Y^+)}{\expop[\ell(U^k)]}\overset{j\to\infty}{\longrightarrow}0.\label{eq:divergencef2vblock}
\end{align}

\subsection{Code Construction}

Define an overflow threshold 
\begin{align}
n\geq k\expop[\ell(U)].\label{eq:threshhold}
\end{align}
This threshold divides the set $\mathcal{C}^k$ into two parts,
\begin{align}
\mathcal{C}_\leq&:=\{c\in\mathcal{C}^k\colon \ell(c)\leq n\}\\
\mathcal{C}_>&:=\{c\in\mathcal{C}^k\colon \ell(c)> n\}
\end{align}
where we write $\mathcal{C}_\leq$ and $\mathcal{C}_>$ without the super-script $^k$ for notational convenience. We define the encoder as follows.
\begin{align}
f_n\colon\mathcal{C}^k\to&\{0,1\}^n\nonumber\\
c\mapsto&f_n(c)=\begin{cases}
cY^{n-\ell(c)}&\text{if }c\in\mathcal{C}_\leq\\
c_1^{n}&\text{if }c\in\mathcal{C}_>
\end{cases}\label{eq:encoder}
\end{align}
where $Y^{n-\ell(c)}$ is a vector of $n-\ell(c)$ random variables that are iid according to $P_Y$ and where $c_1^n$ is $c$ truncated to its first $n$ entries. The mapping $f_n$ defines a random variable $V=f_n(U)$ that takes values in $\{0,1\}^n$. We define
\begin{align}
\mathcal{V}_\leq := f_n(\mathcal{C}_\leq),\quad \mathcal{V}_> := f_n(\mathcal{C}_>).
\end{align}
The support of $V$ can now be written as
\begin{align}
\supp V = \mathcal{V}_\leq \cup \mathcal{V}_>\subseteq \{0,1\}^n.
\end{align}

 Note that $f_n$ is a random mapping. Note further that $f_n\circ h^k$ is a b2b mapping that maps $m=jk$ bits to $n$ bits, i.e.,
\begin{align}
f_n\circ h^k\colon \{0,1\}^m\to \{0,1\}^n.
\end{align}

\subsection{The Role of $j$ and $k$}

We discuss the intuition behind the encoder just defined. Assume $n$ is fixed and given. Because of \eqref{eq:divergencef2v}, the value of $j$ controls how well the matcher output is matched to $P_Y$. The encoder $f_n$ is in part one-to-many (and thereby invertible), and in part many-to-one (which leads to errors when decoding). For a fixed $j$, choosing $k$ small decreases the many-to-one part and thereby the probability of error, but it also decreases the matching rate $jk/n$. Thus $k$ parameterizes a trade-off between probability of error and matching rate. We make this precise in the next section and we illustrate the trade-off for an example in Sec.~\ref{sec:numerical}.
\section{$\epsilon$-Error Matching: Analysis}\label{sec:analysis}
\subsection{Informational Divergence}
\begin{proposition}\label{prop:divergencebound}
The I-divergence per bit achieved by an $\epsilon$-error b2b matcher is upper-bounded by the I-divergence per bit achieved by the internal f2v matcher , i.e.,
\begin{align}
\frac{\kl(P_V\Vert P_Y^n)}{n}\leq \frac{\kl(P_U\Vert P^+_Y)}{\expop[\ell(U)]}.
\end{align}
\end{proposition}
\begin{IEEEproof}
The I-divergence can be written as
\begin{align}
\kl(P_V\Vert P_Y^n)&=\sum_{v\in\mathcal{V}_\leq}P_V(v)\log_2\frac{P_V(v)}{P^n_Y(v)}\nonumber\\
&\qquad+\sum_{v\in\mathcal{V}_>}P_V(v)\log_2\frac{P_V(v)}{P^n_Y(v)}.\label{eq:klf2f}
\end{align}
We write the first sum as 
\begin{align}
&\sum_{v\in\mathcal{V}_\leq}P_V(v)\log_2\frac{P_V(v)}{P^n_Y(v)}\nonumber\\
&=\hspace{-0.25cm}\sum_{\stackrel{c\in\mathcal{C}_\leq}{y\in\{0,1\}^{n-\ell(c)}}}2^{-m}P^{n-\ell(c)}_Y(y)\log_2\frac{2^{-m}P^{n-\ell(c)}_Y(y)}{P^{\ell(c)}_Y(c)P^{n-\ell(c)}_Y(y)}\\
&=\sum_{c\in\mathcal{C}_\leq}2^{-m}\log_2\frac{2^{-m}}{P^{\ell(c)}_Y(c)}\Bigl[\sum_{y\in\{0,1\}^{n-\ell(c)}}P^{n-\ell(c)}_Y(y)\Bigr]\\
&=\sum_{c\in\mathcal{C}_\leq}2^{-m}\log_2\frac{2^{-m}}{P^{\ell(c)}_Y(c)}.\label{eq:firstsum}
\end{align}
The second sum in \eqref{eq:klf2f} can be bounded as
\begin{align}
&\sum_{v\in\mathcal{V}_>}P_V(v)\log_2\frac{P_V(v)}{P^n_Y(v)}\nonumber\\
&=\sum_{v\in\mathcal{V}_>}\Bigl[\sum_{c\in\mathcal{C}\colon c_1^n=v}2^{-m}\Bigr]\log_2\frac{\sum_{c\in\mathcal{C}\colon c_1^n=v}2^{-m}}{P^n_Y(c_1^n)}\\
&\overset{(a)}{\leq}\sum_{v\in\mathcal{V}_>}\Bigl[\sum_{c\in\mathcal{C}\colon c_1^n=v}2^{-m}\Bigr]\log_2\frac{\sum_{c\in\mathcal{C}\colon c_1^n=v}2^{-m}}{\sum_{c\in\mathcal{C}\colon c_1^n=v} P^{\ell(c)}_Y(c)}\\
&\overset{(b)}{\leq}\sum_{v\in\mathcal{V}_>}\sum_{c\in\mathcal{C}\colon c_1^n=v}2^{-m}\log_2\frac{2^{-m}}{P^{\ell(c)}_Y(c)}\\
&=\sum_{c\in\mathcal{C}_>}2^{-m}\log_2\frac{2^{-m}}{P^{\ell(c)}_Y(c)}\label{eq:secondsum}
\end{align}
where we have an inequality in (a) because the prefix-free extensions of $c_1^n$ in $\mathcal{C}$ may not form a complete tree. The inequality in (b) follows by the log-sum inequality \cite[Theorem~A.4]{kramer2012information}.
Using \eqref{eq:firstsum} and \eqref{eq:secondsum} in \eqref{eq:klf2f}, we get
\begin{align}
&\kl(P_V\Vert P_Y^n)\nonumber\\
&\leq \sum_{c\in\mathcal{C}_\leq}2^{-m}\log_2\frac{2^{-m}}{P^{\ell(c)}_Y(c)}+\sum_{c\in\mathcal{C}_>}2^{-m}\log_2\frac{2^{-m}}{P^{\ell(c)}_Y(c)}\\
&=\sum_{c\in\mathcal{C}_\leq\cup\mathcal{C}_>}2^{-m}\log_2\frac{2^{-m}}{P^{\ell(c)}_Y(c)}\\
&=\kl(P_U^k\Vert P_Y^+).\label{eq:divergencebound}
\end{align}
Thus, we have
\begin{align}
\frac{\kl(P_V\Vert P_Y^n)}{n}\overset{(a)}{\leq}&\frac{\kl(P_U^k\Vert P_Y^+)}{n}\\
\overset{(b)}{\leq}&\frac{\kl(P_U^k\Vert P_Y^+)}{\expop[\ell(U^k)]}\\
= &\frac{k\kl(P_U\Vert P^+_Y)}{k\expop[\ell(U)]}\\
= &\frac{\kl(P_U\Vert P^+_Y)}{\expop[\ell(U)]}
\end{align}
where (a) follows by \eqref{eq:divergencebound} and where (b) follows by \eqref{eq:threshhold}. This concludes the proof of the proposition.
\end{IEEEproof}
The statement of Prop.~\ref{prop:divergencebound} can be intuitively explained by the definition \eqref{eq:encoder} of our encoder. For input strings causing underflow, we randomly generate the missing bits according to the target distribution $P_Y$. Thus, these bits do not contribute to the I-divergence. For the input strings causing overflow, we use a many-to-one mapping by truncation. This can only decrease the I-divergence because of the convexity of I-divergence.

\subsection{Probability of Error}
We use letter typicality on $B^j$. Assume $b^{jk}\in\mathcal{T}^k_\epsilon(B^j)$. By the typical average lemma \cite[p. 26]{ElGamal2011},
\begin{align}
\frac{\ell(b^{jk})}{k}\leq (1+\epsilon)\expop[\ell(B^j)]\label{eq:typicallength}
\end{align}
We choose
\begin{align}
n=(1+\epsilon)k\expop[\ell(U)].\label{eq:typicalthreshhold}
\end{align}
We define the decoder as
\begin{align}
\varphi_n\colon \mathcal{V}_\leq\cup\mathcal{V}_>\to &\{0,1\}^{jk}\\
v\mapsto\hat{b}^{jk}=\varphi_n(v)=&\begin{cases}
(f_n\circ h^k)^{-1}(v)&\text{if }v\in\mathcal{V}_\leq\\
\text{error}&\text{if }v\in\mathcal{V}_>.
\end{cases}
\end{align}
By \eqref{eq:typicallength} and \eqref{eq:typicalthreshhold}, an error can only occur if $B^{jk}\notin\mathcal{T}^k_\epsilon(B^j)$, i.e., if the binary sequence to be encoded is not typical. The probability of error is thus bounded by
\begin{align}
\probop[B^{jk}\neq \hat{B}^{jk}]\leq&\probop[B^{jk}\notin\mathcal{T}^k_\epsilon(B^j)]\\
\leq &\delta_\epsilon(P_{B^j},k).
\end{align}
This probability can be made arbitrarily small by choosing $k$ large.
\subsection{Rate}
The rate is
\begin{align}
\frac{m}{n}&=\frac{kj}{(1+\epsilon)k\expop[\ell(U)]}\\
&=\frac{j}{(1+\epsilon)\expop[\ell(U)]}\\
&=\frac{1}{1+\epsilon}\cdot\frac{\entop(U)}{\expop[\ell(U)]}.
\end{align}
In \cite{bocherer2013rooted}, the following implication is shown.
\begin{align}
\frac{\kl(P_U\Vert P_Y^+)}{\expop[\ell(U)]}\overset{j\to\infty}{\longrightarrow}0\Rightarrow \left|\frac{\entop(U)}{\expop[\ell(U)]}-\entop(P_Y)\right|\overset{j\to\infty}{\longrightarrow}0.
\end{align}
Thus, because of \eqref{eq:divergencef2v},
\begin{align}
\frac{m}{n}=&\frac{1}{1+\epsilon}\cdot\frac{\entop(U)}{\expop[\ell(U)]}\\
\overset{j\to\infty}{\to}&\frac{1}{1+\epsilon}\entop(P_Y).
\end{align}
The value of $\epsilon$ can be chosen arbitrarily small. This shows that our matcher can asymptotically achieve the maximum entropy rate of $\entop(P_Y)$.
\section{Numerical Results}
\label{sec:numerical}
\begin{figure}[t]
\centering
\footnotesize
\def\svgwidth{1.0\columnwidth}
\executeiffilenewer{figures/rateVsError.svg}{figures/rateVsError.pdf}%
{inkscape -z -D --file=figures/rateVsError.svg %
--export-pdf=figures/rateVsError.pdf --export-latex}%
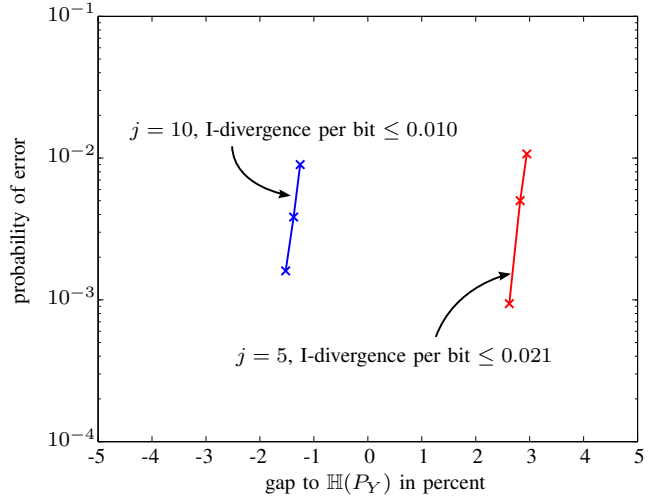%

\caption{The trade-off between rate and probability of error is shown for the target distribution $P_Y$ with $P_Y(0)=0.2$, $P_(1)=0.8$. The output block length of the $\epsilon$-error b2b matcher is $n=58\,320$. Internally, a f2v matcher with $j=5$ (red curve) and $j=10$ (blue curve) is used.}
\label{fig:ratevserror}
\vspace{-0.5cm}
\end{figure}
We illustrate the trade-off between I-divergence, rate, and probability of error of an $\epsilon$-error b2b matcher by an example. We consider the target distribution $P_Y$ with $P_Y(0)=0.2$ and $P_(1)=0.8$. The overflow threshold of the b2b matcher is $n=58\,320$. In Fig.~\ref{fig:ratevserror} the trade-off between rate and probability of error is displayed for internal f2v matchers with $j=5$ (red curve) and $j=10$ (blue curve). In horizontal direction, the gap between the rate and the target entropy $\entop(P_Y)$ is displayed. In vertical direction, the probability of error is shown. As we can see, for $j=10$, we need to use a lower rate to achieve the same probabilities of error as for $j=5$. However, via Prop.~\ref{prop:divergencebound}, we can see from Fig.~\ref{fig:oto} that the matcher with $j=10$ achieves a smaller I-divergence per bit than the matcher with $j=5$.
\section*{Acknowledgment}
Georg B\"ocherer thanks Gerhard Kramer for inspiring discussions, which contributed to the development of this work.
\bibliographystyle{IEEEtran}
\normalsize
\bibliography{IEEEabrv,confs-jrnls,references}

\end{document}

%% file: preamble.tex
\usepackage[colorlinks]{hyperref}
\usepackage{makeidx}
\makeindex

\usepackage{amsmath}
\usepackage{amsthm}
\usepackage{amssymb}
\usepackage{bm}
\usepackage{xspace}
\usepackage{xcolor}
\usepackage{graphicx}
\usepackage{url}
\usepackage{framed}
\usepackage{float}
\usepackage{rotating}
\usepackage{verbatim}
\usepackage{listings}
\usepackage{lscape}

\usepackage{pgfplots}
\usepackage{pgf}
\usepackage{tikz}
\usetikzlibrary{arrows,shapes.misc,chains,scopes}
\pgfplotsset{compat = newest}
\usepackage{pgfplotstable}
\usepackage{booktabs}
\usepackage{colortbl}

\newcommand{\executeiffilenewer}[3]{%
\ifnum\pdfstrcmp{\pdffilemoddate{#1}}%
{\pdffilemoddate{#2}}>0%
{\immediate\write18{#3}}\fi%
}

\graphicspath{{figures/}}

\theoremstyle{plain}
\newtheorem{proposition}{Proposition}

\newcounter{algocount}
\newcounter{examplecount}

\DeclareMathOperator{\probop}{Pr}

\DeclareMathOperator{\expop}{\mathbb{E}}
\DeclareMathOperator{\entop}{\mathbb{H}}
\DeclareMathOperator{\miop}{\mathbb{I}}
\DeclareMathOperator{\kl}{\mathbb{D}}

\DeclareMathOperator*{\minimize}{minimize}

%% file: figures/fixedVsVariable.pdf_tex

\begingroup
  \makeatletter
  \providecommand\color[2][]{%
    \errmessage{(Inkscape) Color is used for the text in Inkscape, but the package 'color.sty' is not loaded}
    \renewcommand\color[2][]{}%
  }
  \providecommand\transparent[1]{%
    \errmessage{(Inkscape) Transparency is used (non-zero) for the text in Inkscape, but the package 'transparent.sty' is not loaded}
    \renewcommand\transparent[1]{}%
  }
  \providecommand\rotatebox[2]{#2}
  \ifx\svgwidth\undefined
    \setlength{\unitlength}{544.35700936pt}
  \else
    \setlength{\unitlength}{\svgwidth}
  \fi
  \global\let\svgwidth\undefined
  \makeatother
  \begin{picture}(1,0.76715379)%
    \put(0,0){\includegraphics[width=\unitlength]{fixedVsVariable.pdf}}%
    \put(0.10750929,0.0218789){\color[rgb]{0,0,0}\makebox(0,0)[lb]{\smash{0}}}%
    \put(0.38077123,0.0218789){\color[rgb]{0,0,0}\makebox(0,0)[lb]{\smash{5}}}%
    \put(0.64897585,0.0218789){\color[rgb]{0,0,0}\makebox(0,0)[lb]{\smash{10}}}%
    \put(0.92237922,0.0218789){\color[rgb]{0,0,0}\makebox(0,0)[lb]{\smash{15}}}%
    \put(0.07085074,0.05392435){\color[rgb]{0,0,0}\makebox(0,0)[lb]{\smash{0}}}%
    \put(0.04528496,0.14638151){\color[rgb]{0,0,0}\makebox(0,0)[lb]{\smash{0.05}}}%
    \put(0.05554287,0.23885697){\color[rgb]{0,0,0}\makebox(0,0)[lb]{\smash{0.1}}}%
    \put(0.04528496,0.33131408){\color[rgb]{0,0,0}\makebox(0,0)[lb]{\smash{0.15}}}%
    \put(0.05554287,0.42362423){\color[rgb]{0,0,0}\makebox(0,0)[lb]{\smash{0.2}}}%
    \put(0.04528496,0.51608134){\color[rgb]{0,0,0}\makebox(0,0)[lb]{\smash{0.25}}}%
    \put(0.05554287,0.60855683){\color[rgb]{0,0,0}\makebox(0,0)[lb]{\smash{0.3}}}%
    \put(0.04528496,0.70086699){\color[rgb]{0,0,0}\makebox(0,0)[lb]{\smash{0.35}}}%
    \put(0.5147082,0){\color[rgb]{0,0,0}\makebox(0,0)[lb]{\smash{m}}}%
    \put(0.01339186,0.25339327){\color[rgb]{0,0,0}\rotatebox{90}{\makebox(0,0)[lb]{\smash{I-divergence per bit}}}}%
  \end{picture}%
\endgroup

%% file: figures/rateVsError.pdf_tex

\begingroup
  \makeatletter
  \providecommand\color[2][]{%
    \errmessage{(Inkscape) Color is used for the text in Inkscape, but the package 'color.sty' is not loaded}
    \renewcommand\color[2][]{}%
  }
  \providecommand\transparent[1]{%
    \errmessage{(Inkscape) Transparency is used (non-zero) for the text in Inkscape, but the package 'transparent.sty' is not loaded}
    \renewcommand\transparent[1]{}%
  }
  \providecommand\rotatebox[2]{#2}
  \ifx\svgwidth\undefined
    \setlength{\unitlength}{551.15720835pt}
  \else
    \setlength{\unitlength}{\svgwidth}
  \fi
  \global\let\svgwidth\undefined
  \makeatother
  \begin{picture}(1,0.77333248)%
    \put(0,0){\includegraphics[width=\unitlength]{rateVsError.pdf}}%
    \put(0.10793593,0.04407503){\color[rgb]{0,0,0}\makebox(0,0)[lb]{\smash{-5}}}%
    \put(0.18883247,0.04407503){\color[rgb]{0,0,0}\makebox(0,0)[lb]{\smash{-4}}}%
    \put(0.26986147,0.04407503){\color[rgb]{0,0,0}\makebox(0,0)[lb]{\smash{-3}}}%
    \put(0.35076345,0.04407503){\color[rgb]{0,0,0}\makebox(0,0)[lb]{\smash{-2}}}%
    \put(0.43179245,0.04407503){\color[rgb]{0,0,0}\makebox(0,0)[lb]{\smash{-1}}}%
    \put(0.52341725,0.04407503){\color[rgb]{0,0,0}\makebox(0,0)[lb]{\smash{0}}}%
    \put(0.60431927,0.04407503){\color[rgb]{0,0,0}\makebox(0,0)[lb]{\smash{1}}}%
    \put(0.68534824,0.04407503){\color[rgb]{0,0,0}\makebox(0,0)[lb]{\smash{2}}}%
    \put(0.76625025,0.04407503){\color[rgb]{0,0,0}\makebox(0,0)[lb]{\smash{3}}}%
    \put(0.84727922,0.04407503){\color[rgb]{0,0,0}\makebox(0,0)[lb]{\smash{4}}}%
    \put(0.92832637,0.04407503){\color[rgb]{0,0,0}\makebox(0,0)[lb]{\smash{5}}}%
    \put(0.04750953,0.06890289){\color[rgb]{0,0,0}\makebox(0,0)[lb]{\smash{$10^{-4}$}}}%
    \put(0.04750953,0.28194423){\color[rgb]{0,0,0}\makebox(0,0)[lb]{\smash{$10^{-3}$}}}%
    \put(0.04750953,0.49496737){\color[rgb]{0,0,0}\makebox(0,0)[lb]{\smash{$10^{-2}$}}}%
    \put(0.04750953,0.70786353){\color[rgb]{0,0,0}\makebox(0,0)[lb]{\smash{$10^{-1}$}}}%
    \put(0.37418478,0.00395529){\color[rgb]{0,0,0}\makebox(0,0)[lb]{\smash{gap to $\entop(P_Y)$ in percent}}}%
    \put(0.01328106,0.28229125){\color[rgb]{0,0,0}\rotatebox{90}{\makebox(0,0)[lb]{\smash{probability of error}}}}%
    \put(0.17178395,0.53428913){\color[rgb]{0,0,0}\makebox(0,0)[lb]{\smash{$j=10$, I-divergence per bit $\leq 0.010$}}}%
    \put(0.33081994,0.19502693){\color[rgb]{0,0,0}\makebox(0,0)[lb]{\smash{$j=5$, I-divergence per bit $\leq 0.021$}}}%
  \end{picture}%
\endgroup